\RequirePackage{fix-cm}
\documentclass[smallextended]{svjour3}       
\smartqed  
\usepackage{graphicx}
\usepackage{epstopdf}
\begin{document}

\title{Metal-Insulator transition in optical lattice system 
with site-dependent interactions}



\author{Takamitsu Saitou \and Akihisa Koga \and Atsushi Yamamoto}


\institute{T. Saitou and A. Koga\at
Department of Physics, Tokyo Institute of Technology, Tokyo 152-8551, Japan \\
\email{koga@phys.titech.ac.jp}           
\and
A. Yamamoto\at
RIKEN, Advanced Institute for Computational Science, 7-1-26,
Minatojima-minami-machi, Chuo-ku, Kobe, Hyogo 650-0047, Japan
}

\date{Received: date / Accepted: date}

\maketitle

\begin{abstract}
We investigate the half-filled Hubbard model 
with spatially alternating interactions by means of
the two-site dynamical mean-field theory.
It is found that a single Mott transition occurs 
when two kinds of interactions are increased.
This implies that the different interactions are essentially irrelevant 
at the critical point. The nature of the Mott states is also addressed.
\end{abstract}

\section{Introduction}
Ultracold atomic gases have attracted much interest 
since the successful realization of Bose-Einstein condensation 
in a bosonic $^{87}$Rb system~\cite{Rb}. 
Due to its high controllability in the interaction strength, particle number, 
and other parameters, many remarkable phenomena have been observed 
such as the Mott transitions in the bosonic and 
fermionic systems~\cite{Greiner,Joerdens,Schneider}.
Recently, the spatial modulation of the interaction has also been realized 
in the $^{174}$Yb gas system~\cite{Yamazaki}, 
which stimulates further theoretical investigations 
on particle correlations in the ultracold atomic systems. 

One of the interesting questions is how the spatial modulation in the interactions 
affects ground-state properties in the fermionic optical lattice.
Although the optical lattice system with uniform interactions has been studied 
theoretically and experimentally,
the system with alternating interactions, which may be one of the simplest systems, 
has not been discussed so far.
Therefore, it is desired to clarify how the alternating interactions 
affect ground-state properties, in particular, the nature of the Mott transition.

Motivated by background, we consider the infinite-dimensional Hubbard model 
with alternating interactions.
By means of the two-site dynamical mean-field theory (DMFT)~\cite{Potthoff}, 
we discuss how particle correlations affect the metal-insulator transition 
at half-filling. The nature of the Mott phases is also discussed.

The paper is organized as follows.
In Sec. \ref{2}, we introduce the model Hamiltonian and briefly 
summarize our theoretical approach.
We demonstrate how particle correlations affect ground state properties
in Sec. \ref{3}.
A brief summary is given in the last section.

\section{Model and method}\label{2}
We study the Hubbard model with alternating interactions, as
\begin{eqnarray}
H=-t\sum_{\langle i,j\rangle\sigma}c_{i \sigma }^{\dagger}c_{j \sigma}
+\sum_{\alpha,i\in\alpha} U_\alpha \left(n_{i\uparrow}n_{i\downarrow}
-\frac{1}{2}\left[n_{i\uparrow}+n_{i\downarrow}\right]\right),
\label{Hami}\end{eqnarray}
where $c_{i\sigma}^\dag (c_{i\sigma})$ creates (annihilates) a fermion 
at the $i$-th site with spin $\sigma(=\uparrow, \downarrow)$. 
$t$ is the hopping integral, and 
$U_\alpha$ is the site-dependent onsite interaction 
in the sublattice $\alpha(=A, B)$.
At half-filling, the systems with $(U_A, U_B)$, $(U_B, U_A)$, $(-U_A, -U_B)$ 
and $(-U_B, -U_A)$ are identified since the Hamiltonian is invariant 
under the particle-hole transformations~\cite{Shiba} as
$c_{i\uparrow}\rightarrow c_{i\uparrow}$ and 
$c_{i\downarrow}\rightarrow (-1)^ic_{i\downarrow}^\dag$.

To investigate the Hubbard model eq. (\ref{Hami}),
we make use of DMFT~\cite{Metzner,Muller,Pruschke95,Georges96,Kotliar04}.
which has successfully been applied to various strongly correlated electron systems.
In the framework of DMFT, the lattice model is mapped to
 an effective impurity  model, 
where local particle correlations are taken into account precisely. 
The Green function  for the original lattice system is then obtained 
via self-consistent equations imposed on the impurity problem.

In DMFT,
the Green function in the lattice system~\cite{Chitra} is given as,

\begin{equation}
G_{\sigma}({\bf k},z)=
\left[
\begin{array}{cc}
z +\mu_{A} -\Sigma _{A \sigma}(z) & -\varepsilon_{\bf k} \\
 -\varepsilon_{\bf k}  & z +\mu_{B} -\Sigma _{B \sigma}(z) 
\end{array}
\right]^{-1},
\end{equation}
where $\mu_\alpha$ and $\Sigma_{\alpha}(z)$ are the chemical potential and 
the self-energy for the $\alpha$-th sublattice, and 
$\epsilon_{\bf k}$ is the dispersion relation for the bare band.
In terms of the density of states (DOS) $\rho_0 (x)$,
the local Green function is expressed as,
\begin{eqnarray}
G_{\alpha\sigma}(z)&=&\int_{-\infty}^{\infty}\frac{\eta_{\bar{\alpha}\sigma}(z)}{\eta_{A\sigma}(z)\eta_{B \sigma}(z)-\varepsilon^2}\rho_0(\varepsilon)d\varepsilon,
\end{eqnarray}
where 
\begin{eqnarray}
\eta_{\alpha\sigma}(z)=z+\mu_\alpha-\Sigma_{\alpha\sigma}(z).
\end{eqnarray}
In the following, we use the semicircular DOS with the half bandwidth $D$,
$\rho_0(x)=\frac{2}{\pi D}\sqrt{1-(x/D)^2}$,
which corresponds to the infinite-coordination Bethe lattice.
Then the self-consistency condition for the sublattice $\alpha$ is 
represented as,
\begin{eqnarray}
{\cal G}_{0\alpha\sigma}(z)&=&z+\mu_\alpha
-\left(\frac{D}{2}\right)^2 G_{\bar{\alpha}\sigma}(z),
\end{eqnarray}
where ${\cal G}_{0\alpha}$ is the non-interacting Green function of 
the effective impurity model for the sublattice $\alpha$.

There are various numerical methods to solve the effective impurity problem.
To discuss the Mott transitions in the Hubbard model at half-filling,
we use here the two-site DMFT method~\cite{Potthoff},
which provides us with a transparent view of
the phase transitions~\cite{Ono,DegKoga}. 
In this scheme, we introduce a specific Anderson impurity model 
for the sublattice $\alpha$, which is connected to only one host site, as
\begin{eqnarray}
H^{imp}_\alpha&=&\varepsilon^{(d)}_\alpha \sum_\sigma 
d^{\dag}_{\alpha\sigma} d_{\alpha\sigma}+
\varepsilon^{(f)}_\alpha\sum_{\sigma }f^{\dag}_{\alpha\sigma}f_{\alpha\sigma}
+V_\alpha \sum_\sigma (d^{\dag}_{\alpha\sigma} f_{\alpha\sigma} 
+ {\rm H.c.})\nonumber\\
&+&U_\alpha \left(n_{f\alpha\uparrow}n_{f\alpha\downarrow}-\frac{1}{2}\left[n_{f\alpha\uparrow}+n_{f\alpha\downarrow}\right]\right),
\label{AndHami}
\end{eqnarray}
where $d_{\alpha\sigma} (f_{\alpha\sigma})$ annihilates a fermion with spin $\sigma$ 
in the effective bath (localized band) and 
$n_{f\alpha\sigma}=f_{\alpha\sigma}^\dag f_{\alpha\sigma}$.
Note that the effective parameters in
the impurity model such as the spectrum of host particles
$\varepsilon_\alpha^{(d)}$, the energy level of localized band
$\varepsilon_\alpha^{(f)}$ and 
the hybridization $V_\alpha$, should be determined selfconsistently
so that the obtained results properly reproduce
the original lattice problem. 
When the system is half-filling,
$\varepsilon^{(d)}_\alpha=\varepsilon^{(f)}_\alpha=0$.

This simple method provides sensible results as far
as low-energy properties are concerned.
Namely, for low-energy excitations around the Fermi
surface, the Green function for $f$-fermions may be approximated
by a single pole as
\begin{eqnarray}
G_{\alpha\sigma}(z) \sim \frac{w_{\alpha}}{z},
\end{eqnarray}
where the residue $w_{\alpha\sigma}$ corresponds to 
the quasi-particle weight for the sublattice $\alpha$. 
Therefore, the self-consistency condition is simplified as
\begin{eqnarray}
V_\alpha &=& \frac{D}{2}\sqrt{w_{\bar{\alpha}}}.\label{eq}
\end{eqnarray}
By estimating the quasi-particle weight for each Anderson impurity model,
we proceed to perform the DMFT iterations.
In the following, we discuss the metal-insulator transition
in the system with alternating interactions.

\section{Results}\label{3}

Let us discuss the stability of a metallic
state in the Hubbard model with spatially-modulated interactions. 
By making use of the two-site DMFT method, 
we calculate the quasi-particle weight for each sublattice,
which is defined as
\begin{eqnarray}
z_\alpha&=&\left(1-\frac{d {\rm Re}\Sigma_\alpha(\omega)}{d\omega}\right)^{-1}.
\end{eqnarray}
\begin{figure}[htb]
\begin{center}
\includegraphics[width=11cm]{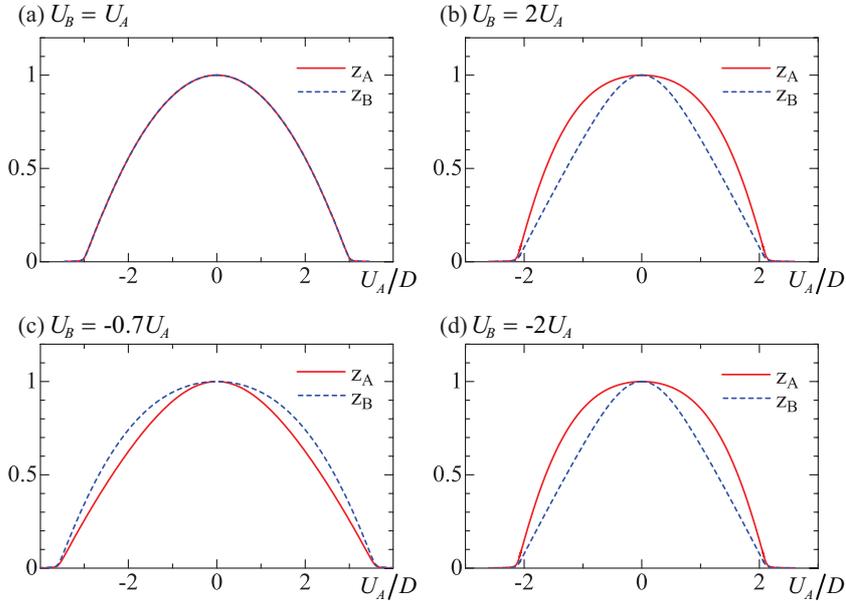}
\caption{Solid (dashed) line represent the quasi-particle weight $z_A (z_B)$
for the $A (B)$ sublattice obtained 
in the systems with fixed ratios $U_B/U_A=1.0$ (a), 
2.0 (b), -0.7 (c) and -2.0 (d).
}
\label{fig1}
\end{center}
\end{figure}
The obtained results for fixed ratios $U_B/U_A=1.0, 2.0, -0.7$ and $-2.0$
are shown in Fig. \ref{fig1}.
When $U_B=U_A$, the system is reduced to the Hubbard model 
with the homogeneous interactions.
As the interactions are introduced, the quasi-particle weights are 
decreased, as shown in Fig. \ref{fig1} (a).
Finally, the quasi-particle weights vanish, where
the Mott transition occurs at $U_A=3D$ in the repulsive case and
the pairing transition occurs at $U_A=-3D$ in the attractive case.
These are consistent with other numerical studies
such as the numerical renormalization group $(U_A=2.94D)$~\cite{Bulla} 
and the exact diagonalization $(U_A=-2.98D)$~\cite{Capone}.

When the spatial dependence of interactions is introduced,
the quasi-particle weights depend on the sublattices,
as shown in Figs. \ref{fig1} (b), (c) and (d).
Namely, the strong renormalization appears in the sublattice with a larger
local interaction.
Nevertheless, the quasi-particle weights vanish simultaneously 
in the strong coupling region.
This suggests that particle correlations induce a single phase transition.
To clarify the nature of the single phase transition,
we focus on low energy properties around the Fermi level.
When the heavy metallic state in the vicinity of the critical point is considered,
the quasi-particle weight in the effective Anderson model eq. (\ref{AndHami})
can be expanded in $V_\alpha$~\cite{LDMFT}, as
\begin{eqnarray}
w_{\alpha\sigma}&=&\left(\frac{6V_\alpha}{U_\alpha}\right)^2.
\end{eqnarray}
In this case, the effective hybridization for the other sublattice is comparable
since it is given as
$V_{\bar{\alpha}}=\frac{3D}{U_\alpha}V_\alpha$.
Therefore, the fermions in both sublattices 
are simultaneously renormalized and different interactions are irrelevant 
around the critical points in this approximation.
The critical interactions for the transition are then given as 
\begin{eqnarray}
U_A U_B &=& \left(3D\right)^2.
\end{eqnarray}
\begin{figure}[htb]
\begin{center}
\includegraphics[width=8cm]{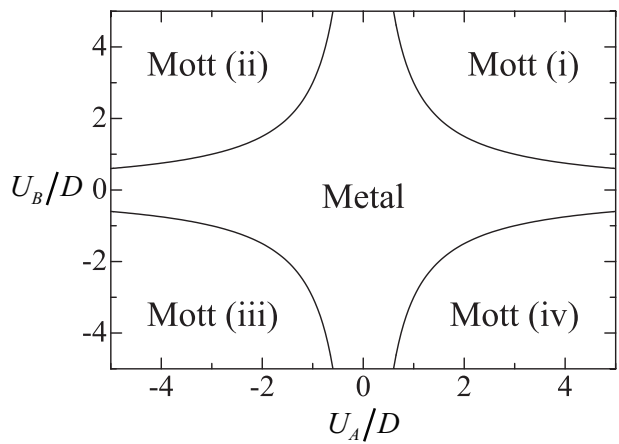}
\caption{Ground state phase diagram for the Hubbard model with spatially-modulated
interactions. }
\label{PD}
\end{center}
\end{figure}
The phase diagram is shown in Fig. \ref{PD}. 
When the interaction is weak, the normal metallic state is realized. 
On the other hand, when the interactions in both sublattices are large,
there exist four kinds of the Mott states, which are denoted as 
the Mott states (i), (ii), (iii), and (iv), as shown in Fig. \ref{PD}.
We find that the Mott state (i) [(iii)] 
in the repulsive (attractive) Hubbard model is widely stabilized 
away from the symmetric limit $(U_A=U_B)$.
In addition, the Mott states (ii) and (iv) are also stabilized
in the system where
interactions are alternately arranged repulsive and attractive.

To clarify the nature of these Mott states, we also calculate 
the double occupancy $D_i(=\langle n_{i\uparrow}n_{i\downarrow}\rangle)$.
\begin{figure}[htb]
\begin{center}
\includegraphics[width=11cm]{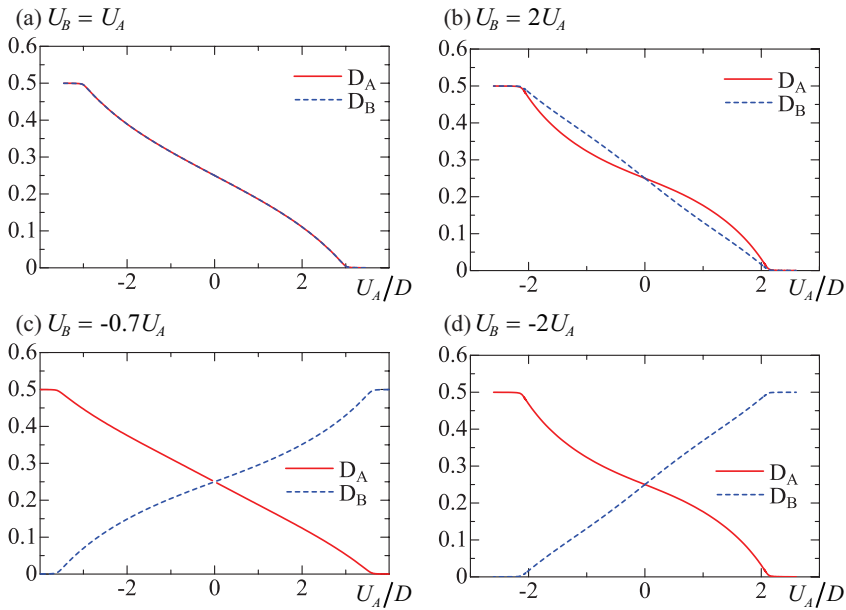}
\caption{Solid (dashed) line represent the double occupancy $D_A (D_B)$
for the $A (B)$ sublattice obtained 
in the systems with fixed ratios $U_B/U_A=1.0$ (a), 
2.0 (b), -0.7 (c) and -2.0 (d).}
\label{fig2}
\end{center}
\end{figure}
In the non-interacting case $(U_A=U_B=0)$, the quantity is quarter
since the empty, single occupied and doubly occupied states
are equally realized at each site.
Introducing the interaction, 
the quantity for each sublattice varies, as shown in Fig. \ref{fig2}.
When $U_B=2U_A$, the double occupancy for the sublattice $B$ rapidly decreases
since the local Coulomb interaction is large.
Finally these quantities for both sublattices vanish simultaneously 
at the Mott transition point.
On the other hand, the double occupancy approaches half 
as the attractive interaction increases.
In this case, the pairing transition occurs at $U_A=-2.12D$ $(U_B=2U_A)$, 
where the single occupied states
are forbidden and the empty or doubly occupied state is realized at each site.
It is also found that 
when the repulsive and attractive interactions alternate in the system $(U_AU_B<0)$,
the Mott and pairing transitions occur simultaneously 
in the corresponding sublattices,
as shown in Figs. \ref{fig2} (c) and (d).

In this paper, we have clarified the nature of the Mott transitions
in the system with alternating interactions, by restricting our discussions
to the paramagnetic case.
Although this condition may be relevant for the optical lattice experiments,
some ordered states are naively expected at very low temperatures.
Therefore, it is necessary to clarify how 
the magnetically ordered state or the superfluid state
is stabilized in the system only with repulsive or attractive interactions
$(U_AU_B>0)$.
On the other hand, the Mott transition should be realized in the system 
with repulsive and attractive interactions $(U_A U_B<0)$ since
any ordered states are hard to be stabilized.
In the case, the hole doping effect is one of the important problems since 
the local particle density varies in the optical lattice system 
with a confining potential.
Furthermore, it is expected that particle correlations induce the commensurability 
in the doped system~\cite{Koga-OSMT,Inaba-OSMT}, which is now under consideration.

\section{Summary}
We have investigated the half-filled Hubbard model 
with spatially-modulated interactions by means of
the two-site dynamical mean-field theory.
It has been clarified that a single Mott transition occurs 
when two kinds of the interactions are increased.
We have determined the ground state phase diagram and 
have discussed the nature of these Mott states.

\begin{acknowledgements}
This work was partly supported by the Grant-in-Aid for Scientific Research 
20740194 (A.K.) and 
the Global COE Program ``Nanoscience and Quantum Physics" from 
the Ministry of Education, Culture, Sports, Science and Technology (MEXT) 
of Japan. 
\end{acknowledgements}



\end{document}